\documentclass[pra,showpacs,showkeys,preprintnumbers,amsmath,amssymb,superscriptaddress,twocolumn]{revtex4-1}
\usepackage[english]{babel}
\usepackage{makeidx} 
\usepackage{graphicx} 
\usepackage{dcolumn} 
\usepackage{array} 
\usepackage{amssymb} 
\usepackage{amsmath}
\usepackage{textcomp}
\usepackage{multirow}
\usepackage{subfigure}
\usepackage{eucal}
\usepackage{mathrsfs}
\usepackage[all]{xy}
\usepackage{epstopdf}
\usepackage{lipsum}
\usepackage{color}
\usepackage{float} 
\usepackage{amsmath} 
\usepackage{amsfonts}
\usepackage{bm}
\begin{document}
\newcommand{\Acal}{\mathcal{A}}
\newcommand{\Bcal}{\mathcal{B}}
\newcommand{\Ccal}{\mathcal{C}}
\newcommand{\Dcal}{\mathcal{D}}
\newcommand{\Ecal}{\mathcal{E}}
\newcommand{\Fcal}{\mathcal{F}}
\newcommand{\Gcal}{\mathcal{G}}
\newcommand{\Hcal}{\mathcal{H}}
\newcommand{\Ical}{\mathcal{I}}
\newcommand{\Jcal}{\mathcal{J}}
\newcommand{\Kcal}{\mathcal{K}}
\newcommand{\Lcal}{\mathcal{L}}
\newcommand{\Mcal}{\mathcal{M}}
\newcommand{\Ncal}{\mathcal{N}}
\newcommand{\Ocal}{\mathcal{O}}
\newcommand{\Pcal}{\mathcal{P}}
\newcommand{\Qcal}{\mathcal{Q}}
\newcommand{\Rcal}{\mathcal{R}}
\newcommand{\Scal}{\mathcal{S}}
\newcommand{\Tcal}{\mathcal{T}}
\newcommand{\Ucal}{\mathcal{U}}
\newcommand{\Vcal}{\mathcal{V}}
\newcommand{\Wcal}{\mathcal{W}}
\newcommand{\Xcal}{\mathcal{X}}
\newcommand{\Zcal}{\mathcal{Z}}

\newcommand{\Lmath}{\mathbbm{L}}
\newcommand{\Rmath}{\mathbbm{R}}
\newcommand{\Hmath}{\mathbbm{H}}

\newcommand{\dket}[1]{| #1 \rangle\rangle}
\newcommand{\dbra}[1]{\langle\langle #1 |}
\newcommand{\ve}[1]{\langle #1 \rangle}
\newcommand{\dinterpro}[2]{\langle \langle #1 | #2 \rangle \rangle}
\newcommand{\interpro}[2]{\langle #1 | #2 \rangle}

\title{Simulating thermodynamic properties of dinuclear metal complexes using Variational Quantum Algorithms}
\affiliation{Grupo de Informa\c{c}\~{a}o Qu\^{a}ntica e F\'{i}sica Estat\'{i}stica, Centro de Ci\^{e}ncias Exatas e das Tecnologias, Universidade Federal do Oeste da Bahia - Campus Reitor Edgard Santos. Rua Bertioga, 892, Morada Nobre I, 47810-059 Barreiras, Bahia, Brazil.}
\author{Ana Clara das Neves Silva\email{ana.s2823@ufob.edu.br}} 
\affiliation{Grupo de Informa\c{c}\~{a}o Qu\^{a}ntica e F\'{i}sica Estat\'{i}stica, Centro de Ci\^{e}ncias Exatas e das Tecnologias, Universidade Federal do Oeste da Bahia - Campus Reitor Edgard Santos. Rua Bertioga, 892, Morada Nobre I, 47810-059 Barreiras, Bahia, Brazil.}

\author{{Lucas Queiroz Galvão\email{lucas.queiroz@fbter.org.br}}} 
\affiliation{{LAQCC - Latin American Quantum Computing Center. SENAI CIMATEC, Salvador, BA, Brazil.}}
\affiliation{{QuIIN - Quantum Industrial Innovation, Centro de Competência Embrapii Cimatec. SENAI CIMATEC, Av. Orlando Gomes, 1845, Salvador, BA, Brazil CEP 41850-010}}

\author{Clebson Cruz} \email{Corresponding author: Clebson Cruz \\ email: \url{clebson.cruz@ufob.edu.br}}  \affiliation{Grupo de Informa\c{c}\~{a}o Qu\^{a}ntica e F\'{i}sica Estat\'{i}stica, Centro de Ci\^{e}ncias Exatas e das Tecnologias, Universidade Federal do Oeste da Bahia - Campus Reitor Edgard Santos. Rua Bertioga, 892, Morada Nobre I, 47810-059 Barreiras, Bahia, Brazil.}
\date{\today}
 
\begin{abstract}

In this paper, we investigate the use of variational quantum algorithms for simulating the thermodynamic properties of dinuclear metal complexes. Our study highlights the potential of quantum computing to transform advanced simulations and provide insights into the physical behavior of quantum systems. The results demonstrate the effectiveness of variational quantum algorithms in simulating thermal states and exploring the thermodynamic properties of low-dimensional molecular magnetic systems. The findings from this research contribute to broadening our understanding of quantum systems and pave the way for future advancements in materials science through quantum computing.

\end{abstract}

\maketitle

\section{Introduction}

As quantum computing progresses, there is a pressing need for the development of materials to meet the growing technical requirements \cite{Iannaccone2018Quantum,Bova2021Commercial,Bassman2021Simulating,Grimsley2018An}. A thorough comprehension of physical and chemical events at nanoscale scales is necessary to create these materials, as quantum effects are crucial in this context. Nevertheless, standard computational methods frequently lack the ability to handle the intricacy of advanced quantum systems. Quantum computing has been extensively researched in recent decades with the potential to revolutionize complex simulations \cite{Montanaro2015Quantum,Ma2020Quantum,Benenti2001Efficient}. 

In this scenario, quantum computing offers the promise of efficiently solving problems currently intractable for classical computers, making it a highly sought-after technology in the field of materials science \cite{Ma2020Quantum,Bova2021Commercial,Bassman2021Simulating,Grimsley2018An,Parrish2019Quantum}. By harnessing the power of quantum mechanics, researchers aim to develop new materials with tailored properties that can meet the demands of modern technological advancements \cite{Marsden2020Quantum,Ma2020Quantum,Bova2021Commercial,Bassman2021Simulating}. Therefore, quantum computing has the potential to significantly accelerate the development of new materials and advance scientific research in various fields. 

Recently, advancements in quantum processing units (QPUs) have led to the potential for "quantum supremacy," where a QPU performs tasks considered impossible for classical computers \cite{king2024computational}. Part of the efforts to harness the capabilities of current quantum devices have resulted in the development of various Quantum Algorithms with a wide range of applications \cite{Montanaro2015Quantum}. In this regard, the intersection between classical computing and quantum computing gave rise to a class of algorithms that challenge the conventional limits of computing: {Variational Quantum Algorithms} (VQAs) \cite{cerezo2021variational,mcclean2016theory}. These algorithms have emerged as powerful tools for exploring and understanding complex quantum systems, providing valuable insights into their physical behavior and properties. 

Based on the theories embedded in the development of VQAs, some innovative approaches to simulating quantum states have been proposed \cite{verdon2019quantum,Grimsley2018An,Parrish2019Quantum,Tilly2021The,Cerezo2020Variational,Wu2018Variational,Li2019Variational,Zeng2020A,medina2023vqeinspired,Endo2018Variational,sun2021quantum,Francis2020Many,kandala2017hardware,luo2024variational,consiglio2023variational,Holmes2022quantumalgorithms}. These algorithms use a quantum circuit and classical optimization routines to simulate the quantum states of a system, given its Hamiltonian. By leveraging the principles of quantum mechanics, VQAs have the potential to revolutionize various fields, such as chemistry and material science \cite{cerezo2021variational,Parrish2019Quantum,Grimsley2018An,Li2019Variational}. 

In this context, this work aims to verify the application of a Variational Quantum Algorithm in simulating thermal states of a quantum system by exploring the Variational Quantum Thermilizer (VQT) routine \cite{verdon2019quantum,Melo2022Implementation}. The VQT algorithm offers a promising avenue for efficiently simulating thermal states, which are crucial for understanding the behavior of quantum systems at finite temperatures \cite{Powers2021Exploring,sun2021quantum,Melo2022Implementation}. 
Furthermore, using this approach, we show that it is possible to obtain the thermodynamic properties, such as specific heat and magnetic susceptibility, of a dinuclear metal complex by comparing the performed quantum simulations with experimental data of the compound $ NH_4CuPO_4\cdot H_2O$ \cite{KOO2008276,Chakraborty_2014}. 

The results obtained in this study highlight the effectiveness of VQAs in accurately predicting the thermodynamic properties of solids. By bringing to light the potential of this algorithm, the present study expands our understanding of Variational Quantum Algorithms and offers substantial contributions to their future practical application. This perspective not only represents a significant achievement in the current context but also lays the foundation for future advances in the emerging field of predicting the physical properties of quantum materials through quantum computing, outlining a promising path for the continued evolution and designing of advanced materials with specific properties. 

\section{Variational Quantum Algorithms for Simulating Quantum States at Finite Temperature}\label{sec2}

With advances in quantum computing, the opportunity arises to explore its applications in problems that demand high computational capacity. A notable example of the application of quantum algorithms is quantum simulation, where so-called Variational Quantum Algorithms (VQAs) \cite{cerezo2021variational,mcclean2016theory} play a fundamental role in the simulation of atomic and molecular systems \cite{kandala2017hardware,watergroundstate}. VQAs are based on a parameterized quantum circuit and utilize a hybrid optimization approach that involves intersections between classical and quantum computing. This intersection aims to find efficient solutions to quantum problems. The versatility and efficiency of VQAs enable diverse applications, ranging from quantum simulations \cite{kandala2017hardware,watergroundstate,peruzzo2014variational,higgott2019variational,Parrish2019Quantum,luo2024variational} and optimizations \cite{farhi2014quantum,li2020quantum} to machine learning \cite{verdon2017quantum,biamonte2017quantum,verdon2018universal}.

The complexity and details of the structure of a Quantum Variational Algorithm (VQA) vary depending on the nature of the problem to be solved \cite{cerezo2021variational}. However, most of these algorithms share fundamental elements such as a cost function (or loss function), a parameterized ansatz, and an optimization loop that optimizes the ansatz \cite{cerezo2021variational,verdon2019quantum,Tilly2021The}. In this context, when facing a specific problem, it is crucial to establish a cost function $\mathrm{C}$ that, based on some variational principle, encodes the solution \cite{verdon2019quantum,cerezo2021variational}. Thus, one can propose an ansatz structured as a quantum operation that depends on a set of parameters $\phi$. {The ansatz is an important part of variational algorithms because it sets up the initial quantum state so that it matches the wave function of the circuit: $U(\phi)\vert{\Psi_0}\rangle= \vert{\Psi (\phi)}\rangle$. Therefore, the task is to optimize the ansatz to find the parameters that minimize the associated cost, following the variational principle \cite{Cerezo2020Variational}: }
\begin{equation}
    E_{min} \leq E(\phi) = \langle{\Psi (\phi) }\vert \Hcal \vert{\Psi(\phi)}\rangle~,
\end{equation}
which guarantees a good approximation of the ground state of the system's Hamiltonian $\Hcal$. The structure of the cost function and the details of the ansatz depend on the specific problem being addressed. However, in many cases, the cost is a function of a set of observables $\{O_k\}$, a set of input states $\{\rho_k\}$, and the parameterized unitary operator $U(\phi)$ \cite{cerezo2021variational}. Thus, this cost function is usually expressed as
\begin{equation}
        C(\phi) = \sum_k f_k\left(tr[O_kU(\phi)\rho_k U^\dagger(\phi)]\right)~.
        \label{eq:cost}
\end{equation}
The importance of VQAs in solving quantum problems lies in the construction of these elements, which is crucial to achieving an effective representation of the desired quantum state \cite{cerezo2021variational}.

The Variational Quantum Eigensolver (VQE) is a commonly used algorithm in quantum computing for determining the ground state of a system based on its Hamiltonian \cite{Tilly2021The,Cerezo2020Variational,Cerezo2020Variational}. This algorithm has proven to be successful, which has led to the development of more comprehensive approaches capable of simulating mixed states and reproducing thermal states of quantum systems \cite{Cerezo2020Variational,kandala2017hardware,Tilly2021The,Parrish2019Quantum,medina2023vqeinspired}. An example of such an approach is the Variational Quantum Thermalizer Algorithm (VQT) \cite{verdon2019quantum}.

VQT is a quantum algorithm that combines machine learning techniques to reproduce thermal states in quantum systems based on a specific Hamiltonian \cite{verdon2019quantum}. This fact is accomplished by approximating the desired thermal state {$\widehat{\sigma}_{\beta}$, given by $e^{-\beta\widehat{H}}/{Z}_{\beta}$, where ${Z}_{\beta}$ is the canonical partition function of the system at temperature $T$}, through a quantum-probabilistic ansatz $\widehat{\rho}_{\theta\phi}$, parameterized by the variables ${\theta, \phi}$  \cite{verdon2019quantum}. Thus, the goal of VQT is to tune these parameters to minimize the cost $\mathcal{L}_{\theta,\phi}$ given by the relative free energy of the ansatz concerning a target Hamiltonian $\widehat{H}$:
\begin{equation}
\mathcal{L}_{\theta,\phi} = \beta \text{tr}(\widehat{\rho}_{\theta\phi}\widehat{H}) -S(\widehat{\rho }_{\theta\phi})~,
\label{custo}
\end{equation}
where $\beta = {1}/{T}$ is a target inverse temperature and $S(\widehat{\rho}_{\theta,\phi})$ is the von Neumann entropy of the ansatz. {This cost function, Eq. (\ref{custo}), can be understood in terms of the quantum relative entropy $ D(\widehat{\rho}_{\theta,\phi} || \widehat{\sigma}_{\beta})$  \cite{verdon2019quantum}, which is defined as: 
\begin{eqnarray}
 D(\widehat{\rho}_{\theta,\phi} || \widehat{\sigma}_{\beta}) &= &-S(\widehat{\rho}_{\theta,\phi}) - \text{tr} (\widehat{\rho}_{\theta,\phi} \log \widehat{\sigma}_{\beta})\nonumber\\ &=& -S(\widehat{\rho}_{\theta,\phi}) + \beta \text{tr}(\widehat{\rho}_{\theta\phi}\widehat{H}) + \log{{Z}_{\beta}}\nonumber\\ &=& \mathcal{L}_{\theta,\phi}  + \log{{Z}_{\beta}}~.   
 \label{eqr}
\end{eqnarray}
Except for the partition function $Z_{\beta}$, which does not depend on the variational parameters $\theta,\phi$ and therefore does not influence the optimization, our cost function, shown in Eq. (\ref{custo}), is directly defined as the quantum relative entropy, as given by Eq. (\ref{eqr}) \cite{verdon2019quantum}}. It quantifies the loss of information that results from incorrectly describing a system that is actually in state \(\hat{\sigma}_{\beta}\) as being in state \(\hat{\rho}_{\theta \phi}\) \cite{murphy2012machine,potts2019introduction}. Thus, by minimizing this cost function, which is achieved by adjusting the ansatz parameters, we can accurately approximate the thermal state of the system.

\begin{figure*}[!] 
  \centering
  \includegraphics[width=\textwidth]{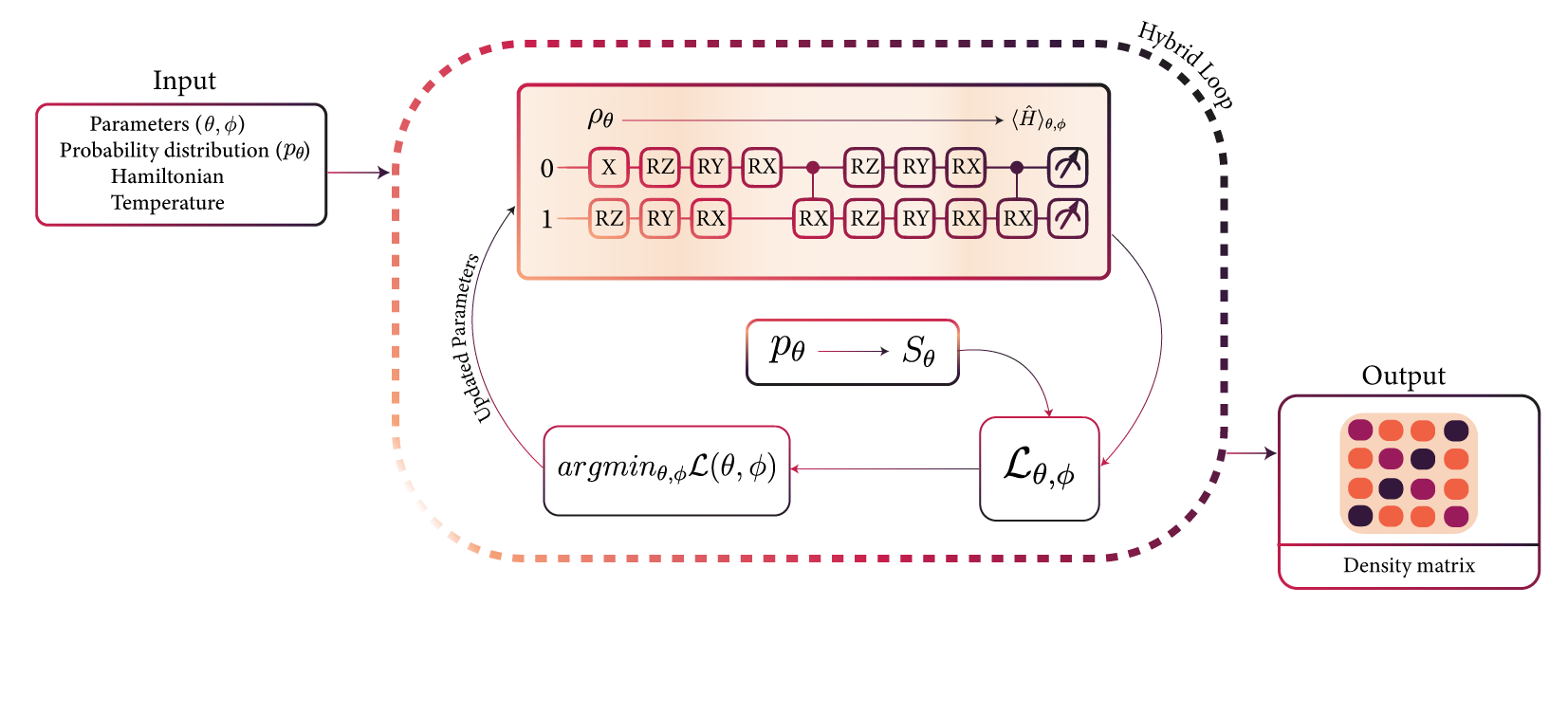} %
  \caption{Schematic representation of the Variational Quantum Thermalizer based Algorithm: Given a Hamiltonian $\hat{H}$, a target inverse temperature, and a parameterized probability distribution $p_{\theta}$, the VQT starts with a mixed state $\rho_ {\theta} $ described by the probability distribution. We generate instances of this probability distribution to obtain individual pure states, which are then passed through a parameterized quantum circuit $U(\phi)$, composed of rotational $RX,~RY,~RZ$ and {controlled-X rotation gate} between the qubits (in our case, 2 qubits). From the results of this circuit, the expectation value of the Hamiltonian about pure states is calculated. Repeating this process several times and averaging these values gives us the expected value of the Hamiltonian concerning $U(\phi)\rho_{\theta}U^{\dagger}(\phi)$. Due to the invariance of the von Neumann entropy $S_{\theta}$ under unitary transformations, it is calculated classically using the parameters $\theta_i$ of the distribution. Once $\langle \hat{H}\rangle_{\theta, \phi}$ and $S_{\theta}$ were obtained, we calculate the cost given by Eq. (\ref{custo}), and the parameters are updated until the relative free energy value is minimum.  At the end of this hybrid loop, using the optimal parameters $\theta,\phi$, we can construct the density matrix of our target mixed state.}
  \label{fig1}
\end{figure*}

In this work, {we used the VQT protocol \cite{verdon2019quantum}} to generate thermal states for a two-qubit system, defined as a dinuclear metal complex with a $d^{9}$ electronic configuration ($s=1/2$) \cite{cruz,cruz2020quantifying,cruz2022,cruz2022quantum}.  Figure \ref{fig1} depicts a schematic representation of the used protocol. The algorithm begins with an input that includes the Hamiltonian model ($\hat{H}$), the desired reciprocal temperature ($1/T$), and the parameterized probability distribution ($p_{\theta}$). 

 {In this regard, the algorithm's first step is to define a mixed state $\rho_{\theta_i}$, for the two-qubit system in its factored form:
\begin{eqnarray}
     \rho_{\theta_i} =  \rho_{\theta_i}^{(1)}\otimes \rho_{\theta_i}^{(2)}~.
\end{eqnarray}
For each single-qubit system, we have
}
{
\begin{eqnarray}
  {\rho_{\theta_i}^{(\alpha)} = p_i(\theta_i)\vert{0}\rangle\langle{0}\vert  + (1 -p_i(\theta_i))\vert{1}\rangle\langle{1}\vert}~,
\end{eqnarray}
where we use the probability distribution $ p_i(\theta_i)$ in order to generate pure states, with $\alpha =\{ 1,2\}$. It is worth highlighting that the advantage of using this factored model is that the number of parameters $\theta_i$ needed to prepare the state scales linearly rather than exponentially with the number of subsystems \cite{verdon2019quantum}.}
Thus, the exact number of parameters depends on the configuration of states within each subsystem \cite{verdon2019quantum}. The pure states are then passed through a parameterized quantum circuit $U(\phi)$, which is made up of rotational $RX,~RY,~RZ$ gates and {controlled-X rotation gates} between the two qubits, as can be seen in  Figure \ref{fig1}.  

 From the results of this circuit, the expectation value of the Hamiltonian about pure states is calculated. {After} repeating this process several times and averaging these values, the algorithm gives us the expected value of the Hamiltonian concerning the operator $U(\phi)\rho_{\theta}U^{\dagger}(\phi)$. Due to the invariance of the von Neumann entropy $S_{\theta}$ under unitary transformations, it can be computed in a classical manner using the distribution parameters $\theta_i$. After obtaining $\langle \hat{H}\rangle_{\theta, \phi}$ and $S_{\theta}$, we calculate the cost using Eq. (\ref{custo}) and update the parameters until the relative free energy value is minimized. At the end of this hybrid loop, we can use the optimal parameters $\theta,\phi$ to construct the density matrix of our desired mixed state.

 For the classical optimization problem, we utilized the Constrained Optimization BY Linear Approximation (COBYLA)  method \cite{powell1994advances}. COBYLA is a numerical optimization technique designed for derivative-free constrained optimization problems where the objective function's derivative is unknown. It involves an iterative approach for derivative-free constrained optimization problems that have been advanced and refined over the years and is widely used in quantum-classical hybrid algorithms \cite{cheng2024quantum,singh2023benchmarking,pellow2021comparison}.  

{In this regard, to implement the quantum algorithm in an ideal quantum processor we used Pennylane's Statevector Simulator \cite{bergholm2022pennylane}. Pennylane is an open-source Python library that offers a flexible platform for simulating materials using quantum circuit simulations, containing advanced methods for VQAs, preparation of thermal states, and evaluation of molecular spectra \cite{Bassman2021Simulating}.  Furthermore, to analyze the algorithm's performance in a more realistic approach, we used the IBM Quantum Experience simulators, \texttt{qasm\_simulator} and \texttt{FakeSheerbrooke} in the IBM Qiskit library \cite{qiskit2024}. The \texttt{qasm\_simulator} in Qiskit is a classical backend for simulating quantum circuits described in a shot-based environment. \texttt{FakeSheerbrooke} is a noisy simulator in Qiskit that mimics the noise characteristics of IBM’s Sherbrooke real quantum computer. It allows users to test quantum algorithms in an environment that closely resembles the real Sherbrooke quantum device, including its realistic noise model.}

{It is important to emphasize that the results presented in this paper were obtained by rebuilding Pennylane’s built-in routine \cite{JackCeroni2020} to meet the specifics of the simulation of metal complex systems. In this regard, we implement a routine to handle the variation of the system's magnetic coupling between the qubits, which was not possible in the original algorithm. This adjustment is fundamental for control and experimentation in our context, as we were able to analyze the thermodynamic properties for a wide range of temperatures and different coupling values ($J$) between the spins.} In the following, we present the assessed thermodynamic properties of the material under investigation and discuss the implications of using quantum computing for future material engineering. We demonstrate the accuracy of the approach discussed in this section in predicting the thermodynamic properties of metal complexes. 

\section{Thermodynamic Properties of a Dinuclear Metal Complex}\label{sec3}

Given a dinuclear metal complex with a Heisenberg dimer structure and an electronic configuration of $d^9$, this system serves as an ideal representation of a two-qubit system (spin $1/2$ dimer). The behavior of this system is governed by the Heisenberg-Dirac-van Vleck Hamiltonian \cite{cruz,cruz2017influence,cruz2020quantifying,cruz2022quantum,cruz2022,Chakraborty_2014}, which can be written in the local basis $\lbrace \vert 00\rangle,\vert 01\rangle,\vert 10\rangle,\vert 11\rangle\rbrace$ as
\begin{eqnarray}
    \mathcal{H}&=& J \vec{S}_A\cdot \vec{S}_B~\nonumber \\
    &=& {\frac{J}{4}} \left[
   \begin{matrix} 1 & 0 & 0 & 0 \\
                      0 & -1 & 2 & 0 \\
                      0 & 2 & -1 & 0 \\
                      0 & 0 & 0 & 1
   \end{matrix} \right]~,
    \label{heisenberg}
\end{eqnarray}
where $J$ is defined as the magnetic coupling constant. 

In this context, at thermal equilibrium, the density matrix of the system of interest can be written in the Gibbs form, $\rho(T)=e^{-\mathcal{H}/k_BT}/Z$, where
\begin{equation}
 Z(T)= e^{-\frac{3J}{4k_BT}} + 3e^{\frac{3J}{4k_BT}}~,
 \label{eq:1}
\end{equation} 
is defined as the partition function. Thus, the  density matrix is written on the local basis as \cite{cruz,cruz2017influence,cruz2022} :
\begin{align}
\rho &=\dfrac{1}{2}\left[
\begin{matrix} 2\varrho_2 & 0 & 0 & {0} \\
0& {\varrho_1 + \varrho_3} & {\varrho_3 - \varrho_1}  & 0\\
0& {\varrho_3 - \varrho_1}  & {\varrho_1 + \varrho_3} & 0\\
0& 0& 0 & 2\varrho_4 
\end{matrix} \right].
\label{eq:2}
\end{align}
Considering the Hamiltonian of the system, Eq. \eqref{heisenberg}, from the spectral decomposition of the density matrix of the system, Eq. \eqref{eq:2}, we obtain the set of eigenvalues $\varrho_{n}$ (population) and its respective eigenvectors $\vert{\varrho_{n}}\rangle$ in terms of a thermodynamic property of the material, namely the molar magnetic susceptibility  \cite{cruz,cruz2017influence,cruz2020quantifying}:
\begin{align}
&\varrho_1 = \varrho_2 = \varrho_3 = \frac{k_BT\chi(J,T)}{2N_Ag^2\mu_B^2} \rightarrow \left\{ 
\begin{aligned} &\vert{\varrho_{1}}\rangle  =\vert 00\rangle \\
&\vert{\varrho_{2}}\rangle  =\frac{1}{\sqrt{2}}\left( \vert 01\rangle + \vert 10\rangle\right) \\
&\vert{\varrho_{3}}\rangle  =\vert 11\rangle \end{aligned}\right.~, \label{eq:07} \\
&\varrho_4= 1 - 3\varrho_1(T) \rightarrow  \vert\varrho_4 \rangle = \frac{1}{\sqrt{2}}\left( \vert 01\rangle {-} \vert 10\rangle\right) ~,
\label{eq:08}
\end{align}
where $g$ is the so-called Land\'{e} factor, $N_A$ is the Avogadro's constant, $\mu_B$ is the Bohr magneton, and $\chi$ is the molar magnetic susceptibility, defided by Bleaney-Bowers  equation\cite{cruz,cruz2017influence,cruz2020quantifying,mario,mario2,brandao2009magnetic,bleaney1952anomalous,yuri,souza,yuri2,Chakraborty_2014}:
\begin{equation}
\chi (J,T) =\frac{2 N_A(g\mu_B)^2}{k_B T}\frac{1}{3+e^{{J}/{k_B T}}}~.
\label{eq:09}
\end{equation}

In the following, the thermodynamic properties of the system of the system can be obtained from Eqs. \eqref{eq:07} and \eqref{eq:08} as $S= - \Sigma_{i=1}^{4} \varrho_i \ln{\varrho_i} $.  Thus, the magnetic entropy of this system can be obtained in terms of the Bleaney-Bowers equation, Eq. \eqref{eq:09}, as:
\begin{widetext}
    \begin{equation}
S(J,T) =  -\frac{k_B T\chi (J,T)}{2 N_A(g\mu_B)^2}  \left\{ e^{\frac{J}{k_BT}}\ln \left[1 - \frac{3k_B T\chi (J,T)}{2 N_A(g\mu_B)^2}\right]  + 3 \ln \left[ \frac{k_B T\chi (J,T)}{2 N_A(g\mu_B)^2}\right]\right\}~,
\label{eq:10}
\end{equation}
\end{widetext}

The relationship between the Bleaney-Bowers magnetic susceptibility and the magnetic entropy provides valuable insights into the behavior of the system at different temperatures. Understanding this connection allows for a deeper analysis of the macroscopic properties of the system and its implications for the overall thermodynamic properties of the material. In this regard, the Bleaney-Bowers magnetic susceptibility is also related to the internal energy of the system, which can be expressed as \cite{yuri,yuri2}:
\begin{equation}
U(J,T) = -3 NJ\left(  \frac{k_B T }{N(g\mu_B)^2}\chi (J,T)-\frac{1}{2}\right)
\end{equation}

As a consequence, the magnetic-specific heat is given by \cite{yuri,yuri2}
\begin{eqnarray}
c_{mag}(T) 
&=& \frac{3 J^2 \chi (J,T)}{8T(g\mu_B)^2}\left[4-\frac{3 k_B T \chi (T)}{2N(g\mu_B)^2}\right]
\label{CalorEspecifico}
\end{eqnarray}

By examining the elements of the system density matrix, researchers can gain a better understanding of the thermodynamic properties of dinuclear metal complexes, such as their molar magnetic susceptibility and magnetic specific heat. Therefore, the simulation of thermal states through VQAs can lead to more accurate predictions of the material's thermodynamic behavior under various conditions, which can lead to future advancements in materials science and technology through the enhancement of material engineering.  

\section{Simulating Thermodynamic Properties Through Quantum Computing}

This section will present the results obtained using the variational algorithm presented in section \ref{sec2} for the described dinuclear spin $1/2$ metal complex. Equation (\ref{heisenberg}) defines the form of the system's Hamiltonian. As can be seen, the matrix entries are parameterized by the system's magnetic coupling constant $J$. {In appendix \ref{appendix} we show the Hamiltonian and density matrix tomography obtained from the statevector simulation.}

In the following, the Hamiltonian can be used to obtain the thermal state, Eq. \eqref{eq:2}. The results reveal the probability distributions $\varrho_{n}$, Eqs. \eqref{eq:07} and \eqref{eq:08}, for each temperature value (see the Appendix \ref{appendix} for more details). As presented in section \ref{sec3}, the simulated thermal state provides valuable insights into the thermodynamic properties of the system, offering a deeper understanding of the influence of the macroscopic properties of the system in its behavior at the quantum level \cite{cruz}.

\begin{figure}[!h]
    \centering
    \includegraphics[width=\columnwidth]{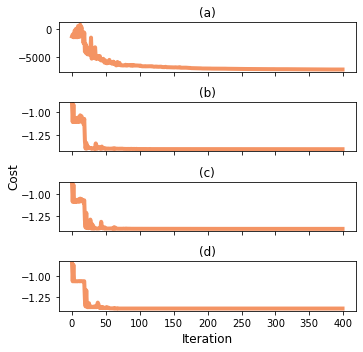}
    \caption{{Performance of the optimization of the cost function given by the Eq. \ref{custo} for different temperature values: (a) $T = 5\times 10^{-4}~K$, (b) $10~K$, (c) $20 ~K$ and (d) $300~K$. For optimization, we used the Constrained Optimization BY Linear Approximation (COBYLA) method. The model was trained for 400 randomly chosen sets of initial parameters with $J/k_B = 1~K$.}}
    \label{otimizacao}
\end{figure}

{We evaluate the thermal states reproduced by the variational algorithm (see section \ref{sec2}) for a set of temperature values, from $T=5\times 10^{-4}~K$ to $350~K$. Figure \ref{otimizacao} shows the performance of minimizing the cost function, given by Eq. (\ref{custo}), during the simulation for high and low-temperature values. In all cases analyzed, the model was trained for 400 randomly chosen initial parameter sets with $J/k_B = 1~K$, and the cost function converges to a minimum value after approximately 200 iterations.}

\begin{figure}[!]
    \centering
    \includegraphics[width=\columnwidth]{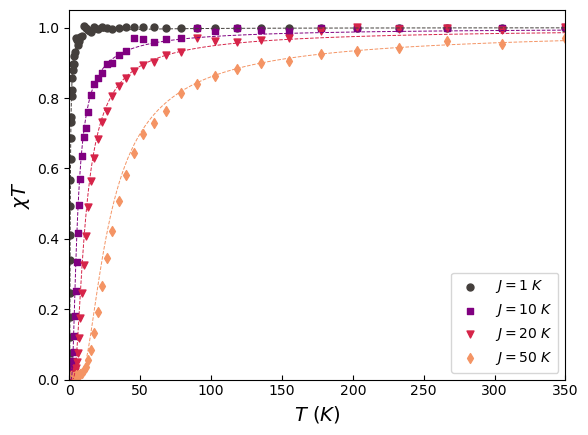}
    \caption{Dimentionless molar magnetic susceptibility, Eq. \eqref{eq:09}, times the temperature for different coupling values $J$. The data were obtained through thermal states reproduced by the variational quantum algorithm presented in section \ref{sec2} considering temperature values in the range of $5\times 10^{-5}~K$ to $350~K$. Dashed line curves represent an analytical simulation of the Bleaney-Bowers equation presented in Eq. \eqref{eq:09}. {The magnetic coupling values were selected to ease the visualization of Schottky’s anomaly of the specific heat.}}
    \label{chi}
\end{figure}

Analyzing the magnetic susceptibility of magnetic systems is a crucial step in comprehending how magnetic properties react to changes in temperature and magnetic coupling ($J$) \cite{mario}. From the simulated thermal state, we can obtain the molar magnetic susceptibility, Eq. \eqref{eq:09}, as a function of the temperature for different magnetic couplings. Figure \ref{chi} shows the product $T\chi(T)$ as a function of the temperature. The behavior of this function is consistent with Curie's Law in the paramagnetic region, where the magnetic orientation of the spins becomes less ordered due to thermal agitation, resulting in the loss of the intrinsic magnetic properties of the system in the paramagnetic transition \cite{mario}. {The magnetic coupling values were selected from 1 K to 50 K to present a wide range of magnetic susceptibility, spanning from low temperatures to near room temperature, approaching the paramagnetic region.} 

\begin{figure}[!]
    \centering
    \includegraphics[width=\columnwidth]{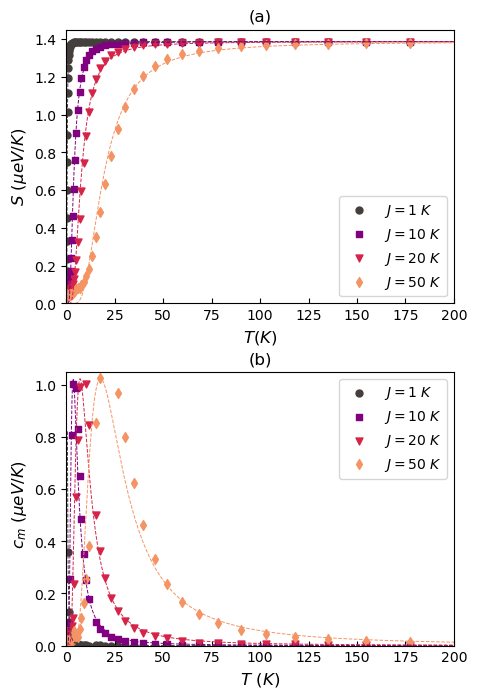}
    \caption{(a) Entropy of the Heisenberg dimer as a temperature function for different coupling values $J$. The data were obtained through thermal states reproduced by VQT (Variational Quantum Thermalizer Algorithm) considering temperature values in the range of $5\times 10^{-5}~K$ to $200~K$. (b) Magnetic specific heat as a function of temperature for different values of magnetic coupling $J$. The data points were derived from entropy data. The curves with dashed lines refer to the theoretical model for the simulated Hamiltonian, Eq. (\ref{heisenberg}). }
    \label{entropiaCE}
\end{figure}

In the following, the entropy of the system can be calculated as $S=-\mbox{Tr}[\rho\ln(\rho)]$.  The graph in Figure \ref{entropiaCE} (a) shows how the entropy of the system changes with temperature for different magnetic coupling values. We also created specific heat curves by examining the relationship between entropy and specific heat, as shown in Figure \ref{entropiaCE} (b). The peaks in the specific heat curves indicate sudden changes in the system's ability to absorb thermal energy, revealing high specific heat values at low temperatures. This behavior reproduces the famous Schottky anomaly at low temperatures \cite{mario}. 

The results obtained using the variational quantum algorithm are in complete agreement with the theoretical expectations for the presented quantities \cite{mario}, demonstrating the accuracy and efficiency of using quantum computing to study the thermodynamic properties of materials. These results highlight the effectiveness of variational quantum algorithms in reproducing efficient approximations of the macroscopic properties of quantum systems. Additionally, the successful application of quantum computing in this study showcases its potential for revolutionizing the field of material science. The ability to accurately predict thermodynamic properties using variational quantum algorithms opens up new possibilities for designing and discovering novel materials with desired characteristics. 

\section{Experimental Comparison: Thermodynamic properties of $ NH_4CuPO_4\cdot H_2O$}

In order to validate our simulation, we conduct a study of the 
$ NH_4CuPO_4\cdot H_2O$ compound \cite{KOO2008276,Chakraborty_2014,nawa2012single,pujana1998synthesis}, a dinuclear Cu(II) metal complex using the proposed variational quantum algorithm.  This material appears as a model in scientific literature to study quantum correlations and behaviors in condensed matter systems \cite{Chakraborty_2014}. In this regard, the molar magnetic susceptibility was investigated in the temperature range of 2 K to 100 K, showing antiferromagnetic correlations in the system \cite{KOO2008276,nawa2012single,pujana1998synthesis}. The experimental data were analyzed using an isolated Heisenberg dimer model, allowing the determination of the exchange coupling constant $J$ \cite{pujana1998synthesis}. In addition, specific heat measurements were conducted throughout a temperature range of 2 K to 100 K, yielding valuable information about the thermodynamic properties of the material at low temperatures \cite{pujana1998synthesis,Chakraborty_2014}.

\begin{figure}[!]
    \centering
    \includegraphics[width=\columnwidth]{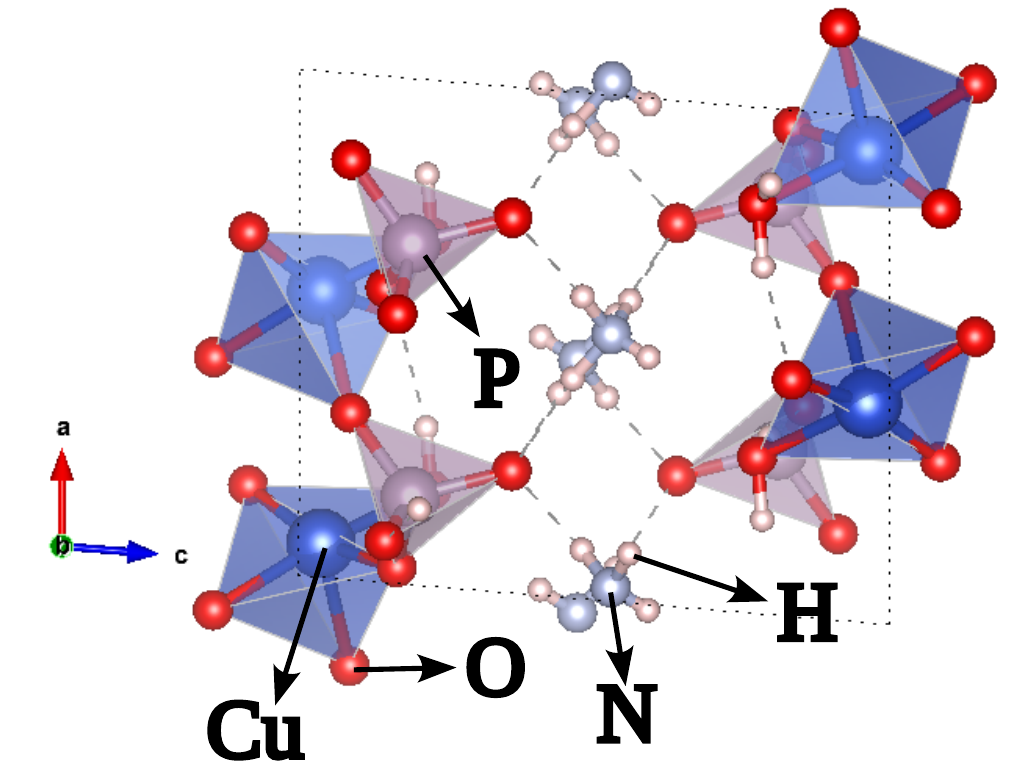}
    \caption{Structure of a single unit cell of the molecule $NH_4CuPO_4\cdot H_2O$. Layers of copper phosphate sheets are linked together by $NH_4^{+}$ cations. The layers are composed of square pyramids of $CuO_5$, which are linked together by phosphate tetrahedra. The Cu(II) ion forms bonds with phosphate oxygens and a water molecule, resulting in the formation of a distinctive layered structure. The crystal structure of this compound was obtained free of charge from the Cambridge Crystallographic Data Centre (CCDC) \cite{ccdc}. The structure was visualized using the Visualisation for Electronic Structural Analysis (VESTA) software \cite{momma2011vesta}.}
    \label{structure}
\end{figure}

Figure \ref{structure} depicts the crystal structure of the $NH_4CuPO_4\cdot H_2O$ compound \cite{ccdc}.  The copper phosphate sheets in the compound are connected by ammonium cations ($NH_4^{+}$). The layers consist of centrosymmetric dimers composed of deformed square pyramids of $CuO_5$ that share edges. These dimers are joined by phosphate tetrahedra with shared corners. Consequently, every metallic center is connected to three phosphate oxygens and one water molecule, creating the foundation of the square pyramid. Meanwhile, the phosphate oxygen associated with the symmetry occupies the axial position \cite{pujana1998synthesis}. Additional details regarding this material's synthesis and crystallographic characterization can be found in the following references: \cite{KOO2008276, nawa2012single, pujana1998synthesis}.

\begin{figure}[!]
    \centering
    \includegraphics[width=\columnwidth]{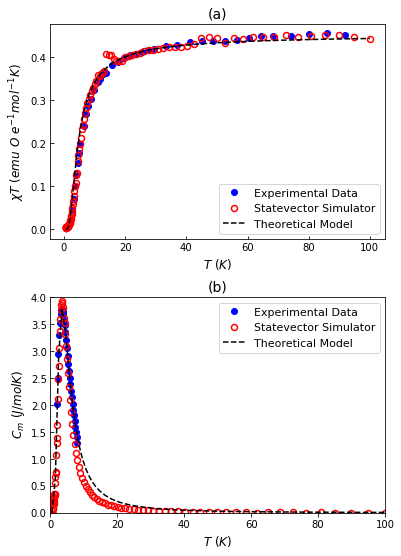}
    \caption{(a) Magnetic susceptibility and (b) Magnetic specific heat for $ NH_4CuPO_4\cdot H_2O$ as a function of temperature. The unfilled red points were obtained using the Variational Quantum Thermalizer Algorithm (VQT) for $J_{VQA}/k_B = 4.2 ~K$, compared to the experimental data (filled blue points)  $J_{exp}/k_B = 5 ~K$. The dashed line represents the theoretical models described in Eqs. \eqref{eq:09} and \eqref{CalorEspecifico}.}
    \label{expEH}
\end{figure}

Therefore, from the magnetic structure of the $NH_4CuPO_4\cdot H_2O$, it can be described as a dinuclear metal complex with metallic centers in a $d^{9}$ electronic configuration, providing a nearly ideal realization of a Heisenberg spin $1/2$ dimer model \cite{Chakraborty_2014,nawa2012single,pujana1998synthesis}. Thus, from the experimental results obtained in the literature for this compound, we simulate the thermal state that would most closely approximate the experimental data for molar magnetic susceptibility and specific heat \cite{newville2016lmfit}.
Through this comparison, we can evaluate the accuracy and effectiveness of the variational quantum algorithm in predicting the thermal properties of the compound based on experimental data.
Figure \ref{expEH} shows the comparison between the experimental data \cite{Chakraborty_2014,nawa2012single,pujana1998synthesis} (unfilled blue points) and the simulated data from the proposed variational quantum algorithm (filled red points).

As can be seen, the proposed variational quantum algorithm accurately reproduces the experimental data for molar magnetic susceptibility and specific heat. The value of the magnetic coupling parameter, obtained from the variational quantum algorithm, that best fit the experimental data was $J_{VQA}/k_B = 4.2 ~K$. This result is in good agreement with the fit of the experimental data $J_{exp}/k_B = 5 ~K$ reported in the literature \cite{Chakraborty_2014,pujana1998synthesis}. Additionally, the specific heat data obtained from the simulation showed a peak of the Schottky anomaly (as described in sections \ref{sec3}) at around $3.5~K$, which is consistent with the reported experimental data \cite{Chakraborty_2014}. The close agreement between the simulated and experimental results indicates the potential of quantum algorithms in material science research. This suggests that the quantum algorithm used in the simulation is effective in predicting the behavior of magnetic materials. 

{However, it is worth noting that the results, even when simulated in an ideal environment, show slight discrepancies with theoretical and experimental results. In this regard, it is important to highlight some of the inherent limitations of variational algorithms: the gradient function in classical optimization aims to find the better parameters $\phi$ that minimize the cost function $C$ (eq. \ref{eq:cost}), but the convergence of these parameters is not always exact. Some effects must be taken into account, such as barren plateaus noise \cite{larocca2024review}. This effect refers to flat regions (or plateaus) in the cost function optimization landscape of these algorithms, where the gradient of the cost function is very small or almost zero \cite{wang2021noise, uvarov2021barren}. The deeper and larger the circuit, the more likely these noises are \cite{kulshrestha2022beinit}. On the other hand, some alternatives to accelerate variational quantum eigensolvers involve integrating advanced Bell measurement techniques \cite{PhysRevResearch.6.013205}. These methods aim to improve efficiency, thereby reducing computational time, which can facilitate advancements in quantum chemistry and material science.}

\subsection{{Realistic noisy simulation}}

{Conversely, in order to analyze the algorithm viability in current quantum devices, the VQT was simulated in shot-based and noisy simulators. For this purpose, weused the IBM Quantum Experience simulators, \texttt{qasm\_simulator} and \texttt{FakeSheerBrooke}. The \texttt{qasm\_simulator} in Qiskit is a classical backend for simulating quantum circuits in a shot-based environment described in QASM format \cite{qiskit2024}.  \texttt{FakeSheerBrooke} is a noisy simulator in Qiskit that mimics the noise characteristics of 127-qubits IBM's Sherbrooke real quantum computer.}  

{To obtain the expected values to be minimized by the optimization process, it is necessary to measure the observable at their respective bases in the quantum circuit. It can be done by adding rotation gates after calculating the state in the ansatz \cite{koch2020fundamentals}. Naturally, qiskit measures values in the computational base, corresponding to the Z base. For the base X, it is necessary to add Hadamard gates on both qubits, while for the base y it is necessary to add a rotation on the x axis with an angle $-\pi/2$. In each circuit we choose 1024 shots to the circuits to measure the expected value of each observable in the system Hamiltonian (\ref{hamiltoniano}). }

\begin{figure}[!]
    \centering
    \includegraphics[width=\columnwidth]{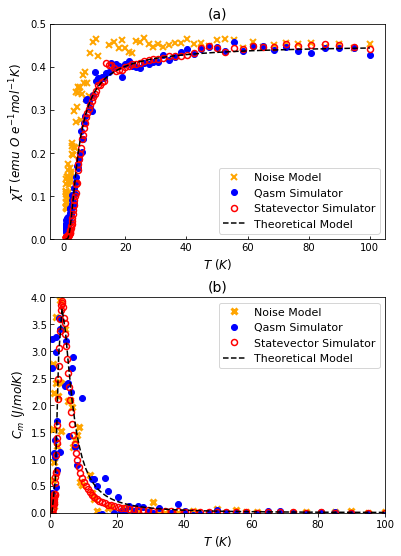}
    \caption{(a) Magnetic susceptibility and (b) Magnetic specific heat for $ NH_4CuPO_4\cdot H_2O$ as a function of temperature {obtained in a single run of the algorithm in a noisy simulator}. The curves show the simulation performance for the shot-based noisy model (orange x-shaped data) in relation to the simulation run in the shot-based qasm\_simulator without noisy (filled blue points) and in the StateVector simulator (unfilled red points) for $J_{VQA}/k_B = 4.2~K$. The dashed line represents the theoretical models described in Eqs. \eqref{eq:09} and \eqref{CalorEspecifico}. In each circuit, we choose 8192 shots and 400 simulation steps in classical optimization. }
    \label{qiskit}
\end{figure}

{In each circuit corresponding to an observable of the Heisenberg Hamiltonian, we choose 8192 shots, which is the default of IBM Qiskit itself. For the process of optimizing the {ansatz} parameters, the same minimization method was used, but adding a limit of 400 iterations at most, taking into account the longer time needed to compile the algorithm in noisy simulators. {To arrive at these configurations, tests were carried out with different numbers of shots and the maximum number of iterations. Figure \ref{qiskit} shows the best test results obtained with the previously specified settings in a single run.}}

{The results show the impact of shot-based and noisy in algorithm in our simulation. When using the \texttt{qasm\_simulator}, it is possible to observes small discrepancies compared to the Statevector simulation. Even without noise, this outcome is expected because the shot-based process generates a probability distribution for each eigenstate based on the number of measurements taken \cite{li2023qasmbench}. The more measurements taken (shots), the more accurate the probability distribution becomes. In our algorithm, we used 8192 shots, which provided a reasonable approximation of the state vector.}

{As expected, the \texttt{FakeSherbrooke} simulator (Noise Model) demonstrated greater discrepancies compared to others. This is a direct result of the fact that the simulator incorporates inherent characteristics of the real Sherbrooke quantum computer. Among these characteristics, those with the greatest influence on the result are: influence of disconnection between qubits \cite{kapit2020entanglement}, decoherence related to the rate of loss of quantum information \cite{shaji2007qubit}, gate error due to a possible failure in its implementation \cite{ayral2021quantum} or noise inherent to measurements in the circuit \cite{resch2021benchmarking}. However, even with these noise effects, it is possible to notice that the outcomes follow the behavior of the theoretical model, with the points being concentrated in regions close to the theoretical curve.}

{In summary, the use of quantum simulators is crucial for the progression of quantum computing, as they enable researchers to foresee and address practical issues prior to executing algorithms on actual hardware . Simulators play a significant role in understanding and mitigating the effects of noise, thereby aiding in the development of more robust and accurate quantum computers. The findings from our study highlight the discrepancies and challenges associated with shot-based and noisy simulations, underscoring the importance of these tools in the pursuit of fault-tolerant quantum computing. As the field advances, ongoing refinement of simulation techniques and noise models will be essential for bridging the gap between theoretical models and practical implementations, particularly in the study of thermodynamic properties of advanced materials using quantum computing.}

\section{Conclusions}

In summary, this work showcases the significant potential of Variational Quantum Algorithms (VQAs) in simulating the thermodynamic properties of advanced materials, particularly focusing on dinuclear spin $1/2$ metal complexes. {Using the Variational Quantum Thermalizer algorithm}, we simulated the evolution of thermal properties with changing temperatures, highlighting the transition from quantum to classical behavior as temperature increases. The simulated results provide valuable insights into the thermodynamic properties of the Heisenberg dimer system, offering a deeper understanding of its behavior at the quantum level. Additionally, we compare the prediction of the molar magnetic susceptibility and the specific heat obtained from the proposed VQA with experimental data reported in the literature for the $ NH_4CuPO_4\cdot H_2O$  dinuclear metal complex. The agreement between the simulated and experimental results validates the accuracy of the VQA method in predicting material properties, demonstrating the promise of quantum algorithms for advancing research in materials science. Therefore, this approach demonstrates the potential of quantum computing in accurately predicting and analyzing material properties, paving the way for more efficient and innovative material design processes.

The intersection of quantum computing and materials science has enormous potential for accelerating the discovery and design of new materials with tailored properties to meet the demands of modern technological advancements. The successful application of VQAs (variational quantum algorithms) to predict thermal properties demonstrates the potential of quantum algorithms to revolutionize complex simulations and drive scientific progress in various fields. These results lay a solid foundation for future research aimed at harnessing the capabilities of quantum computing to predict the thermodynamic properties of advanced materials, thereby opening up exciting possibilities for designing materials with specific functionalities and characteristics. 

\begin{acknowledgments}
The authors thank M. S. Reis for his helpful comments and {Fundação de Amparo à Pesquisa do Estado da Bahia - FAPESB for its financial support (grant number APP0041/2023)}. Lucas Q. Galvão thanks EMBRAPII and the Brazilian Ministry for Science, Technology and Innovation (MCTI) for financial support.
\end{acknowledgments}

\appendix
{
\section{HAMILTONIAN AND DENSITY MATRICES TOMOGRAPHY}\label{appendix}

Figure \ref{hamiltoniano} depicts the tomography of the system Hamiltonian in the computational basis, as defined by Eq. (\ref{heisenberg}). As can be seen, the
matrix entries are parameterized by the system’s magnetic coupling constant $J$.

\begin{figure}[H]
    \centering
    \includegraphics[width=\columnwidth]{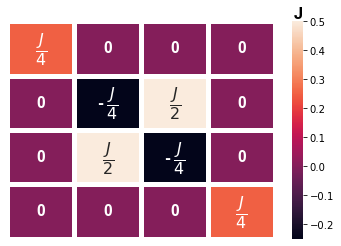}
    \caption{Simulated Hamiltonian matrix, parametrized by
the magnetic coupling constant $J$, in the computational basis.}
    \label{hamiltoniano}
\end{figure}

In the following, the Hamiltonian can be used to obtain the thermal state, Eq. \eqref{eq:2}. The results reveal the probability distributions $\varrho_{n}$, Eqs. \eqref{eq:07} and \eqref{eq:08}, for each temperature value.  Figure \ref{estados} depicts the thermal state
tomography of the system obtained from the variational
quantum algorithm (see section \ref{sec2}).

\begin{figure*}[!]
    \centering
    \includegraphics[scale = 0.6]{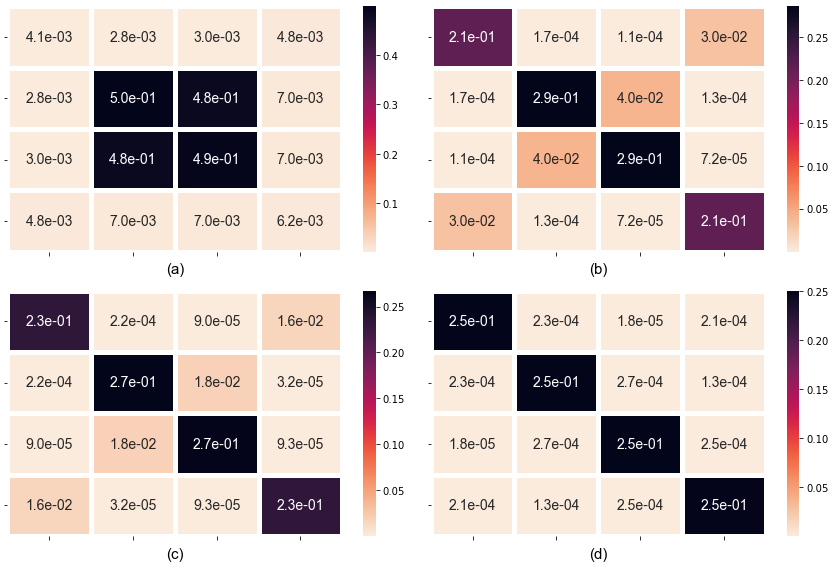}
    \caption{Thermal states obtained in the simulation for different temperature values: (a) thermal state at $T=5\times 10^{-4}~K$; (b) $T=10~K$; (c) $T=20~K$; (d) $T = 300~K$. All these thermal states were reproduced after $400$ simulation steps, with $J/k_B = 1~K$, where $J$ and $k_B$ are the magnetic coupling and the Boltzmann constant, respectively.}
    \label{estados}
\end{figure*}

As can be seen, considering $J/k_B = 1~K$, Figure \ref{estados} (a) shows that at very low temperatures ($T=5\times 10^{-4}~K$), the system exists in the pure singlet ground state $\vert\varrho_4 \rangle = \frac{1}{\sqrt{2}}\left( \vert 01\rangle {-} \vert 10\rangle\right)$, which is a {maximally entangled state}. As we increase the temperature to $10~K$ in Figure \ref{estados} (b), the density matrix approaches a two-qubit X-shaped state \cite{sarandy}, which is characterized by the terms of the secondary diagonal of the matrix. Upon reaching $20~K$ in Figure \ref{estados} (c), the populations increase while the coherence terms decrease. Finally, when the temperature reaches a high enough value of $T = 300~K$, the system tends towards a classical probability distribution, which is characterized by the density matrix in its diagonal form, as shown in Figure \ref{estados} (d). This evolution from an entangled mixed state to a classical probability distribution highlights the transition from quantum to classical behavior as temperature increases. The changes in the density matrix elements reflect the diminishing quantum correlations and the emergence of classical-like features in the system as thermal fluctuations dominate \cite{schlosshauer2019quantum}.

}
\end{document}